\documentclass[twocolumn,english]{article}
\usepackage[T1]{fontenc}
\usepackage[latin9]{inputenc}
\usepackage[a4paper]{geometry}
\usepackage{bm}
\usepackage{amsmath}
\usepackage{amssymb}
\usepackage{graphicx}

\makeatletter
\usepackage{xcolor}            % specify colors by name (e.g. "red")
\definecolor{myred}{rgb}{0.66, 0.15, 0.15}

\usepackage{hyperref}          % clickable URls and cross-references
\hypersetup{
     colorlinks   = true,
     citecolor    = myred,
     urlcolor     = myred,
     urlbordercolor = myred
}
\usepackage{cuted}
\usepackage{cite}

\makeatother

\usepackage{babel}
\begin{document}
\begin{strip}\vspace{-2cm}

\title{\textbf{\Large{}Polaron bubble stabilised by medium-induced three-body
interactions}}
\author{Pascal Naidon\\
{\footnotesize{}Strangeness Nuclear Physics Laboratory, RIKEN Nishina
Centre, Wak{\=o}, 351-0198 Japan. }}
\maketitle

{\color{red} {\bf Important note:} This manuscript contains some errors that will be addressed in a future version. For a proper treatment of the mixed-bubble phenomenon (partial miscibility) described in this manuscript, please refer to the more recent work: \href{https://arxiv.org/abs/2008.05870}{arXiv/2008.05870}. }

\begin{abstract}
Mixing two kinds of particles that repel each other usually results
in either a homogeneous mixture when the repulsion is weak, or a complete
phase separation of the two kinds when their repulsion is too strong.
It is shown however that there is an intermediate regime where the
two kinds can coexist in their ground state as a bubble immersed in
a gas of one kind. Such a situation is obtained by adding heavy repulsive
impurities into a Bose-Einstein condensate. Above a certain strength
of the mutual repulsion, a stable bubble of impurities and bosons
can be formed, resulting from the equilibrium between the interactions
induced by the bosons inside the bubble and the outside pressure from
the surrounding bosons. At some particular strength, the effective
interactions between the impurities consist of only three-body interactions.
Finally, above a critical strength, the bosons are ejected from the
bubble and the impurities collapse into a pure bubble of impurities.
This phenomenon could be observed with an imbalanced mixture of ultracold
atoms of different masses. Moreover, it appears possible to reach
a regime where the impurities form a dense bubble of strongly-interacting
particles.
\end{abstract}
\end{strip}

\section{Introduction}

Mixtures of particles in the quantum regime have been a topic of research
in various fields of physics, starting from experiments on liquid
helium mixtures~\cite{Walters1956,Graf1967}. With the development
of ultracold atom experiments, it has been possible to realise mixtures
of atoms at low temperature and study their properties with full control
over their parameters such as density and interactions. Originally,
the experiments focused on the more stable mixtures of bosonic atoms
with repulsive interactions \cite{Stenger1998,Modugno2002,Papp2008,Lee2018}.
These experiments exhibited the phenomenon of phase separation for
large enough interspecies repulsion, as anticipated by theoretical
works based on the mean-field approximation \cite{Esry1997,Pu1998,Timmermans1998,Ao1998,Trippenbach2000,Pethick2002}.
More recently, there has been an interest in mixtures of bosonic atoms
with attractive interactions~\cite{Cabrera2018,Cheiney2018,Semeghini2018},
originally thought to be unstable, after it was discovered that they
can form self-bound liquid droplets~\cite{Petrov2015}. Even the
properties of a single or a few particles mixed with another kind
of particles constitute a challenging problem for theory. Such impurities
immersed in a medium become quasi-particles known as polarons, and
have been the subject of recent experimental investigations both in
fermionic~\cite{Schirotzek2009,Nascimbene2009,Koschorreck2012,Kohstall2012,Cetina2016,Scazza2017,Yan2019a}
and bosonic~\cite{Catani2012,Park2012,MingGuangHu2016,Jorgensen2016,DeSalvo2017,Camargo2018,Yan2019}
ultracold atomic media, as well as many related theoretical works~\cite{Chevy2006,Prokofev2008,Massignan2014,Schmidt2018,Kalas2006,Cucchietti2006,Tempere2009,Rath2013,Christensen2015,Volosniev2015,Grusdt2015,Vlietinck2015,Ardila2015,Ardila2016,Grusdt2016,Nakano2016,Shchadilova2016,Yoshida2018,Levinsen2017,Parisi2017,Sun2017,Volosniev2017,Grusdt2017,Lampo2017,Loft2017,Sun2017a,Naidon2018a,Camacho-Guardian2018,Guenther2018,Grusdt2018,Pastukhov2018,Lampo2018,Charalambous2019,Mehboudi2019,Kain2018,VanLoon2018,Drescher2019,Mistakidis2019,Watanabe2019,Panochko2019,Takahashi2019,Ichmoukhamedov2019,Boyanovsky2019}.
One compelling aspect of these polarons is their effective interactions
mediated by the medium~\cite{Bardeen1967,Yu2012}.

So far, theoretical works have focused on the pairwise interactions
induced by the medium. In this work, it is shown that the medium may
also induce three-body interactions, which can have a crucial effect
on the macroscopic properties of the impurities. In the case of bosonic
impurities immersed in a Bose-Einstein condensate, a bubble of polarons
can be formed and stabilised by these three-body interactions. This
paper first gives an exact calculation of the induced two-body and
three-body interactions between impurities induced by the surrounding
condensate, in the specific limit of infinitely heavy impurities and
perturbative interactions. In a second step, a simple mean-field theory
is used to characterise the resulting state formed by the impurities.
Finally, a possible implementation with ultracold atoms is discussed.

\section{Mediated interactions}

Let us consider a system of $N_{I}$ particles of mass $M$, referred
to as \emph{impurities}, immersed in a homogeneous gas of $N_{B}$
bosonic particles of mass $m$, referred to as \emph{bosons}. This
system is described in the second-quantisation formalism by the following
Hamiltonian,
\begin{eqnarray}
H & = & \sum_{\bm{k}}\epsilon_{k}b_{\bm{k}}^{\dagger}b_{\bm{k}}+\frac{1}{2V}\sum_{\bm{k},\bm{k}^{\prime},\bm{p}}U_{B}(\bm{p})b_{\bm{k^{\prime}-p}}^{\dagger}b_{\bm{k+p}}^{\dagger}b_{\bm{k}}b_{\bm{k}^{\prime}}\nonumber \\
 &  & +\sum_{\bm{k}}\varepsilon_{k}c_{\bm{k}}^{\dagger}c_{\bm{k}}+\frac{1}{2V}\sum_{\bm{k},\bm{k}^{\prime},\bm{p}}U_{I}(\bm{p})c_{\bm{k^{\prime}-p}}^{\dagger}c_{\bm{k+p}}^{\dagger}c_{\bm{k}}c_{\bm{k}^{\prime}}\nonumber \\
 &  & +\frac{1}{V}\sum_{\bm{k},\bm{k}^{\prime},\bm{p}}U(\bm{p})b_{\bm{k^{\prime}-p}}^{\dagger}c_{\bm{k}+\bm{p}}^{\dagger}c_{\bm{k}}b_{\bm{k}^{\prime}}\label{eq:Hamiltonian}
\end{eqnarray}
where $V$ is the system's volume, $\epsilon_{k}=\frac{\hbar^{2}k^{2}}{2m}$
and $b_{\bm{k}}$ are the kinetic energy and annihilation operator
for a boson with momentum $\bm{k}$, and $\varepsilon_{k}=\frac{\hbar^{2}k^{2}}{2M}$
and $c_{\bm{k}}$ are the kinetic energy and annihilation operator
for an impurity with momentum $\bm{k}$. The potential $U_{B}$ and
$U_{I}$ describe the interaction between two bosons, and between
two impurities, respectively, while the potential $U$ describes the
interactions between a boson and an impurity. The potential $U_{B}$
is assumed to be repulsive to guarantee the stability of the medium
of condensed bosons. Moreover, all potentials $U_{B}$, $U_{I}$,
and $U$ are assumed to be in the perturbative regime satisfying the
Born approximation~\cite{Newton2013}, \emph{i.e.} their respective
scattering lengths $a_{B}$, $a_{I}$, and $a$ can be expanded in
a convergent perturbative series of the form $a=a_{1}+a_{2}+a_{3}+\dots$that
is dominated by the first-order term $a_{1}=\frac{2\mu}{4\pi\hbar^{2}}U(\bm{0})$,
where $\mu$ is the reduced mass of the considered particles. For
interactions $U(\bm{k})$ that become negligible for $k\gtrsim\Lambda$,
where $\Lambda^{-1}$ corresponds to the range of the interaction
(for instance, the nanometre range for neutral atoms), then $a_{2}\approx-a_{1}^{2}\frac{2}{\pi}\Lambda$
and the Born approximation requires that the scattering length $a_{1}$
be much smaller than the range $\Lambda^{-1}$.

In this regime, the bosons can be treated with the Bogoliubov approach~\cite{Bogoliubov1947,Pethick2002},
which consists in applying the following substitution,
\begin{align}
b_{\bm{0}} & \equiv\sqrt{N_{0}}\label{eq:Condensate}\\
b_{\bm{k}} & \equiv u_{k}\beta_{\bm{k}}-v_{k}\beta_{-\bm{k}}^{\dagger}\qquad\text{for }\bm{k}\ne\bm{0}\label{eq:Bogoliubov}
\end{align}
where $N_{0}$ represents the macroscopic number of bosons occupying
the condensate mode $\bm{k}=\bm{0}$, and $\beta_{\bm{k}}$ is the
annihilation operator for bosonic quasi-particles (Bogoliubov quasi-particles),
corresponding to elementary excitations of the bosonic system. The
coefficients $u_{k}$ and $v_{k}$ are chosen to diagonalise the
quadratic form in $\beta$ and $\beta^{\dagger}$ appearing in the
first line of Eq.~(\ref{eq:Hamiltonian}) when the substitution is
applied. The Hamiltonian then reads $H=H^{(\text{F})}+H^{(\text{NF})}$
with
\begin{align}
H^{(\text{F})}= & E_{0}+{\displaystyle \sum_{\bm{k}\ne\bm{0}}E_{k}\beta_{\bm{k}}^{\dagger}\beta_{\bm{k}}}+H^{\prime}\nonumber \\
 & +\sum_{\bm{k}}(\varepsilon_{k}+n_{B}U(\bm{0}))c_{\bm{k}}^{\dagger}c_{\bm{k}}+\sum_{\bm{k}}C_{\bm{k}}(\beta_{\bm{k}}^{\dagger}+\beta_{-\bm{k}})n_{\bm{k}}\nonumber \\
 & +\frac{1}{2V}\sum_{\bm{k},\bm{k}^{\prime},\bm{p}}U_{I}(\bm{p})c_{\bm{k^{\prime}-p}}^{\dagger}c_{\bm{k+p}}^{\dagger}c_{\bm{k}}c_{\bm{k}^{\prime}},\label{eq:HamiltonianH0}\\
H^{(\text{NF})} & ={\displaystyle \frac{1}{V}\sum_{\bm{p}\ne\bm{0}}\sum_{\bm{k}\ne\bm{0},-\bm{p}}\bigg(A_{\bm{k},\bm{p}}\beta_{\bm{k}+\bm{p}}^{\dagger}\beta_{\bm{k}}}\nonumber \\
 & -B_{\bm{k},\bm{p}}\left(\beta_{-\bm{k}-\bm{p}}\beta_{\bm{k}}+\beta_{\bm{k}+\bm{p}}^{\dagger}\beta_{-\bm{k}}^{\dagger}\right)\bigg)n_{\bm{p}},\label{eq:HamiltonianH1}
\end{align}
where $E_{0}\approx\frac{1}{2}U_{B}(\bm{0})Vn_{0}^{2}$ and $E_{k}=\sqrt{\epsilon_{k}(\epsilon_{k}+2n_{0}U_{B}(0))}$
are the Bogoliubov ground-state and excitation energies, and $n_{0}=N_{0}/V$
and $n_{B}=N_{B}/V$ are the condensate and total densities of bosons.
 In the perturbative limit of small $U_{B}$, the term $H^{\prime}$
, which contains orders higher than quadratic in $\beta$ and $\beta^{\dagger}$,
may be neglected. Likewise, the condensate is almost pure, and $n_{0}$
may be approximated by $n_{B}$. In Eqs.~(\ref{eq:HamiltonianH0}-\ref{eq:HamiltonianH1}),
the notation $n_{\bm{k}}\equiv\sum_{\bm{p}}c_{\bm{p}-\bm{k}}^{\dagger}c_{\bm{p}}=n_{-\bm{k}}^{\dagger}$
has been used, and the coefficients $A_{\bm{k},\bm{p}}$, $B_{\bm{k},\bm{p}}$,
and $C_{\bm{k}}$ are given by the following expressions,
\begin{align}
A_{\bm{k},\bm{p}} & \equiv U(\bm{p})\left(u_{\vert\bm{k}+\bm{p}\vert}u_{k}+v_{\vert\bm{k}+\bm{p}\vert}v_{k}\right)\label{eq:CoefA}\\
B_{\bm{k},\bm{p}} & \equiv U(\bm{p})u_{\vert\bm{k}+\bm{p}\vert}v_{k}\label{eq:CoefB}\\
C_{\bm{k}}\equiv & \frac{1}{V}\sqrt{N_{0}}U(\bm{k})(u_{k}-v_{k})\label{eq:CoefC}
\end{align}
where the Bogoliubov amplitudes $u_{k}$ and $v_{k}$ are given by
$u_{k}^{2}=\frac{1}{2}\left(\frac{\epsilon_{k}+n_{0}U_{B}(\bm{0})}{E_{k}}+1\right)$
and $v_{k}^{2}=\frac{1}{2}\left(\frac{\epsilon_{k}+n_{0}U_{B}(\bm{0})}{E_{k}}-1\right)$. 

The Hamiltonian $H^{(\text{F})}$ of Eq.~(\ref{eq:HamiltonianH0})
corresponds to the Fröhlich polaron model~\cite{Froehlich1954,Devreese2009},
in which impurities can move in the bosonic medium and either create
or absorb excitations of the medium through the term proportional
to $C_{\bm{k}}$. It is known~\cite{Devreese2009} that this model
can be solved exactly in the limit of static impurities, i.e. large
mass $M$. Indeed, the following substitution,
\begin{equation}
\beta_{\bm{k}}\equiv\phi_{\bm{k}}-\frac{C_{\bm{k}}}{E_{k}}n_{\bm{k}}.\label{eq:Frohlich}
\end{equation}
formally turns $H^{(\text{F})}$ into
\begin{align}
H^{(\text{F})} & =E_{0}+{\displaystyle \sum_{\bm{k}}E_{k}\phi_{\bm{k}}^{\dagger}\phi_{\bm{k}}}{\displaystyle +\sum_{\bm{k}}(\varepsilon_{k}+E_{P}^{(\text{F})})c_{\bm{k}}^{\dagger}c_{\bm{k}}}\nonumber \\
 & +\frac{1}{2V}\sum_{\bm{k},\bm{k}^{\prime},\bm{p}}\left(U_{I}(\bm{p})+V^{(\text{F})}(\bm{p})\right)c_{\bm{k}+\bm{p}}^{\dagger}c_{\bm{k}^{\prime}-\bm{p}}^{\dagger}c_{\bm{k}^{\prime}}c_{\bm{k}}\label{eq:FrohlichHamiltonian}
\end{align}
 with 
\begin{align}
E_{P}^{(\text{F})} & =n_{B}U(\bm{0})-n_{0}\frac{1}{V}\sum_{\bm{p}}\frac{U(\bm{p})^{2}\epsilon_{p}}{E_{p}^{2}},\label{eq:FrohlichPolaronEnergy}\\
V^{(\text{F})}(\bm{p}) & =-2n_{0}\frac{U(\bm{p})^{2}\epsilon_{p}}{E_{p}^{2}}.\label{eq:FrohlichTwoBodyInteraction}
\end{align}

Equation~(\ref{eq:FrohlichHamiltonian}) formally represents the
Hamiltonian of a system of bosonic quasi-particles (hereafter referred
to as ``Fröhlich quasi-particles'') with annihilation operator $\phi_{\bm{k}}$
and another system of quasi-particles (polarons) with annihilation
operator $c_{\bm{k}}$. One can check that $\phi_{\bm{k}}$ satisfies
the canonical commutation relations $[\phi_{\bm{k}},\phi_{\bm{q}}]=0$
and $[\phi_{\bm{k}},\phi_{\bm{q}}^{\dagger}]=\delta_{\bm{k},\bm{q}}$.
However, the two systems are independent only in the limit of infinite
mass $M$. In this limit, the ground state of the system consists
of the vacuum of Fröhlich quasi-particles and the ground state of
a system of heavy polarons with self-energy $E_{P}^{(\text{F})}$
and effective two-body interaction potential $V^{(\text{F})}$. 

The Hamiltonian $H^{(\text{NF})}$ is proportional to the boson-impurity
interaction $U$, which is assumed to be weak. One may therefore treat
this term as a perturbation to the exact ground state of the Fröhlich
Hamiltonian $H^{(\text{F})}$ obtained in the limit of infinite mass.
Upon making the substitution of Eq.~(\ref{eq:Frohlich}), $H^{(\text{NF})}$
becomes in the vacuum of Fröhlich quasi-particles
\begin{equation}
\langle H^{(\text{NF})}\rangle=\frac{1}{V}\sum_{\bm{p}\ne0}\sum_{\bm{k}\ne0,-\bm{p}}D_{\bm{k},\bm{p}}\frac{C_{\bm{k}}}{E_{\bm{k}}}\frac{C_{\bm{k}+\bm{p}}}{E_{\bm{k}+\bm{p}}}n_{-\bm{k}-\bm{p}}n_{\bm{k}}n_{\bm{p}}.\label{eq:NonFrohlichHamiltonianVacuum}
\end{equation}
with $D_{\bm{k},\bm{p}}=A_{\bm{k},\bm{p}}-B_{\bm{k},\bm{p}}-B_{-\bm{k}-\bm{p},\bm{p}}=U(\bm{p})(u_{k}-v_{k})(u_{\bm{k}+\bm{p}}-v_{\bm{k}+\bm{p}})$.
Using the commutation relations for $c_{\bm{k}}$ and $c_{\bm{k}}^{\dagger}$,
one arrives at the perturbed Hamiltonian:
\begin{align}
\langle H\rangle & =E_{0}+{\displaystyle {\displaystyle \sum_{\bm{k}}(\varepsilon_{k}+E_{P})c_{\bm{k}}^{\dagger}c_{\bm{k}}}}\nonumber \\
 & +\frac{1}{2V}\sum_{\bm{k},\bm{k}^{\prime},\bm{p}}\left(U_{I}(\bm{p})+V(\bm{p})\right)c_{\bm{k}+\bm{p}}^{\dagger}c_{\bm{k}^{\prime}-\bm{p}}^{\dagger}c_{\bm{k}^{\prime}}c_{\bm{k}}\nonumber \\
 & +\frac{1}{6V^{2}}\sum_{\bm{p},\bm{q}}W(\bm{q},\bm{p})\sum_{\bm{k},\bm{k}^{\prime},\bm{k}^{\prime\prime}}c_{\bm{k}+\bm{p}+\bm{q}}^{\dagger}c_{\bm{k}^{\prime}-\bm{q}}^{\dagger}c_{\bm{k}^{\prime\prime}-\bm{p}}^{\dagger}c_{\bm{k}^{\prime\prime}}c_{\bm{k}^{\prime}}c_{\bm{k}}\label{eq:PerturbedHamiltonian}
\end{align}
where $E_{P}=E_{P}^{(\text{F})}+E_{P}^{(\text{NF})}$ and $V=V^{(\text{F})}+V^{(\text{NF})}$
with

\begin{align}
E_{P}^{(\text{NF})} & =\frac{n_{0}}{V^{2}}\sum_{\bm{p},\bm{k}}U(\bm{p})U(\bm{k})U(\bm{s})f_{k}f_{s},\label{eq:PolaronSelfEnergy}\\
V^{(\text{NF})}(\bm{p}) & =2n_{0}\frac{U(\bm{p})}{V}\sum_{\bm{q}}U(\bm{q})U(\bm{s})f_{s}\left(2f_{p}+f_{q}\right),\label{eq:BipolaronPotential}\\
W(\bm{q},\bm{p}) & =2n_{0}U(\bm{p})U(\bm{q})U(\bm{s})\left(f_{q}f_{s}+f_{s}f_{p}+f_{p}f_{q}\right),\label{eq:TripolaronPotential}
\end{align}
and $\bm{s}\equiv\bm{p}+\bm{q}$ and $f_{k}=\epsilon_{k}/E_{k}^{2}$.

\begin{figure}
\hfill{}\includegraphics[width=8.5cm]{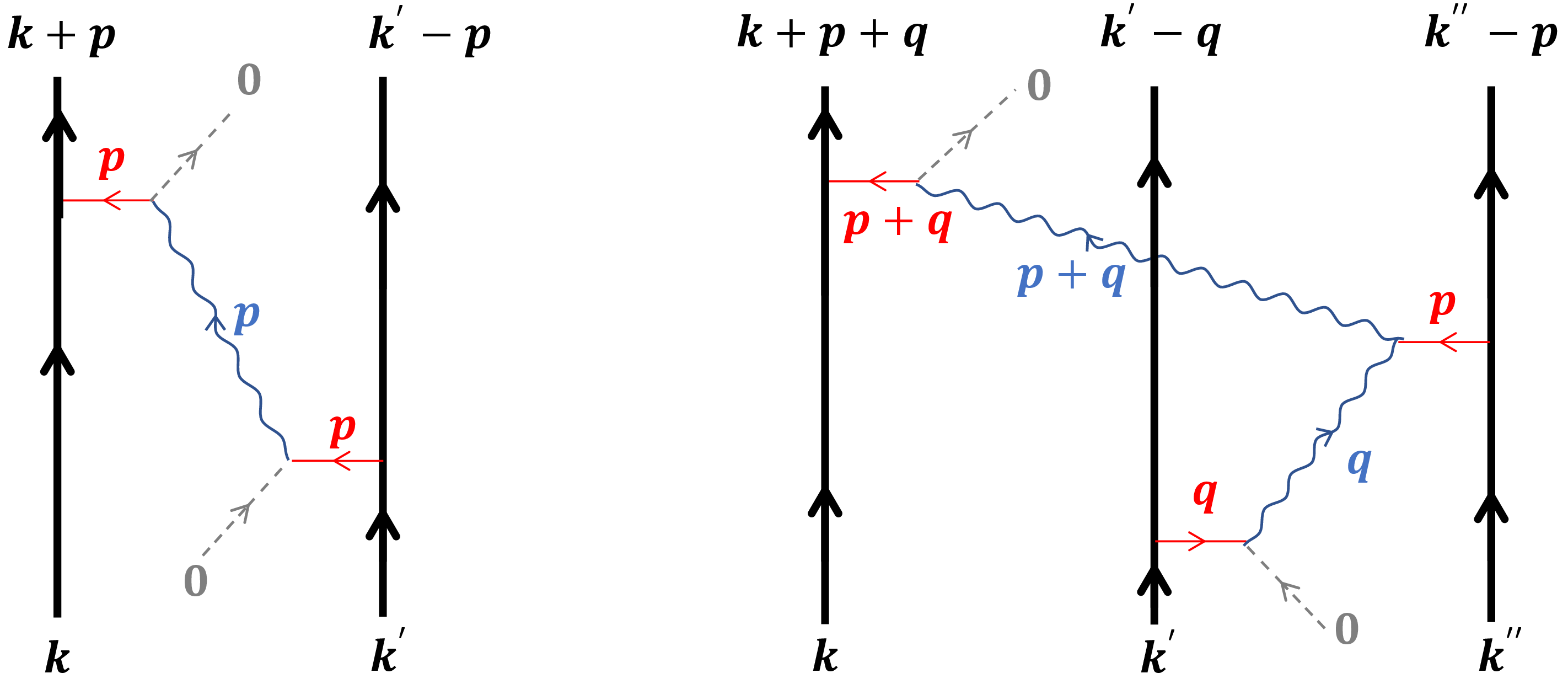}\hfill{}

\caption{\label{fig:diagrams}Leading-order diagrams for the two-body potential
$V$ (left) and three-body potential $W$ (right) given by Eqs.~(\ref{eq:FrohlichTwoBodyInteraction})
and (\ref{eq:TripolaronPotential}). Solid black lines represent impurities,
dashed lines represent condensate bosons, thin red lines represent
boson-impurity interactions, and blue wavy lines represent bosonic
excitations.}

\end{figure}

The expressions for $E_{P}$, $V$, and $W$ are exact in the limit
of a perturbative interaction $U$. The perturbative nature of these
expressions is apparent as they can be easily represented as perturbative
diagrams up to third order in $U$ - see Fig.~\ref{fig:diagrams}.
Interestingly, the non-Fröhlich part of the Hamiltonian not only modifies
the self-energy and induced two-body force between impurities, but
also gives rise to a three-body interaction $W$ between impurities.
This three-body interaction may be interpreted as the leading-order
three-body process in which an impurity creates a Bogoliubov excitation
out of the condensate, which scatters onto a second impurity and is
annhiliated back to the condensate by a third impurity, as shown in
Fig.~\ref{fig:diagrams}.

\begin{figure*}[t]
\centering{}\includegraphics[bb=0bp 0bp 340bp 340bp,width=12.5cm]{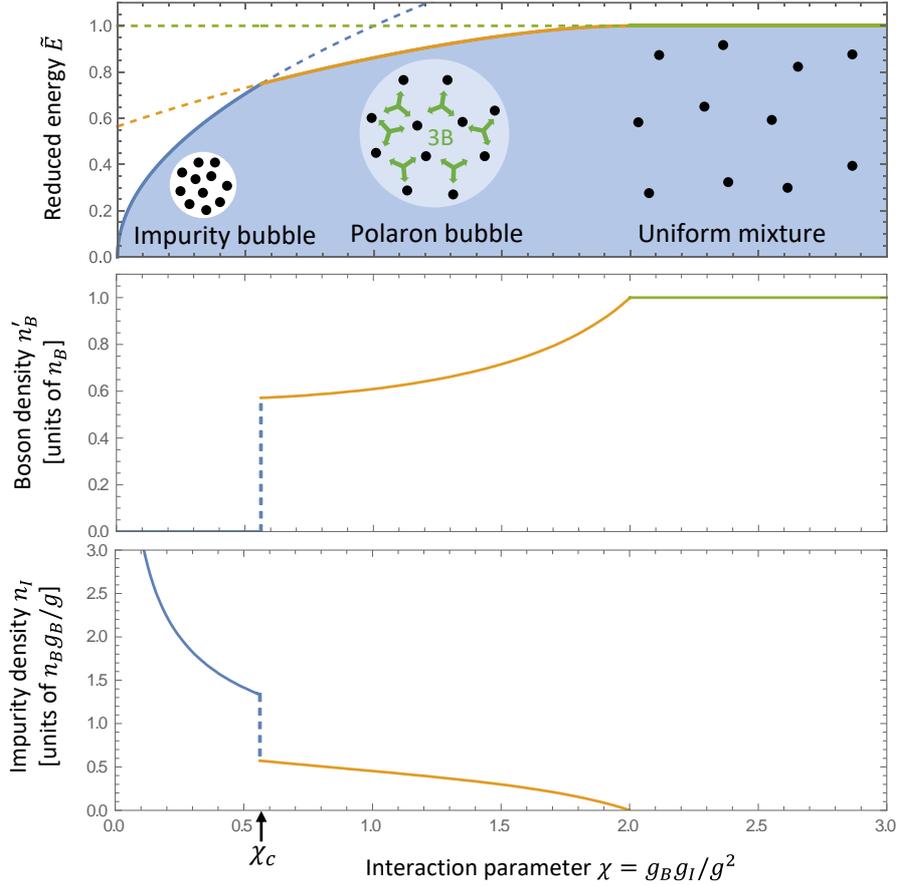}\caption{\label{fig:Scenario}Heavy bosonic impurities immersed in a Bose-Einstein
condensate, as a function of the parameter $\chi=g_{B}g_{I}/g^{2}$
characterising the relative strength of repulsion $g$ between impurities
and bosons, with respect to the self-repulsions $g_{B}$ and $g_{I}$
of bosons and impurities. There are three distinct phases: impurity
bubble, polaron bubble, and uniform mixture. Top panel: energy of
the system, along with a schematic representation of the phases -
the black dots represent impurities and the blue shading represents
the density of the bosonic medium. Middle panel: density of bosons
in-between impurities. Lower panel: density of impurities.}
\end{figure*}

Let us consider an interaction $U(\bm{k})$ that becomes negligible
for $k\gtrsim\Lambda$. If the range $\Lambda^{-1}$ is much smaller
than the coherence length of the bosons $\xi=(8\pi n_{0}a_{B})^{-1/2}$,
then, at distances larger than $\Lambda^{-1}$, one can write $E_{P}$,
$V$, and $W$ in real space as follows:

\begin{equation}
E_{P}=\frac{2\pi\hbar^{2}}{m}n_{B}\!\left(a_{1}\!+\!a_{2}\!+\!a_{3}\!+\!\frac{a_{1}^{2}\!+\!2a_{1}a_{2}}{\tilde{\xi}}\!+\!\frac{a_{1}^{3}}{\tilde{\xi}^{2}}\right)\label{eq:PolaronSelfEnergyPerturbative}
\end{equation}
\begin{equation}
\tilde{V}(\bm{r})=\frac{4\pi\hbar^{2}}{m}n_{B}\Big(-\frac{a_{1}^{2}+2a_{1}a_{2}}{\tilde{\xi}}f(r)+\frac{a_{1}^{3}}{\tilde{\xi}^{2}}f(r)^{2}\Big)\label{eq:BipolaronPotentialPerturbative}
\end{equation}
\begin{equation}
\tilde{W}(\bm{x},\bm{y})=\frac{4\pi\hbar^{2}}{m}n_{B}\frac{a_{1}^{3}}{\tilde{\xi}^{2}}\Big(f(z)f(y)+f(x)f(y)+f(z)f(x)\Big)\label{eq:TripolaronPotentialPerturbative}
\end{equation}
where $\tilde{\xi}=\xi/\sqrt{2}$, $z=\vert\bm{x}-\bm{y}\vert$,
and $f(r)=(r/\tilde{\xi})^{-1}\exp(-r/\tilde{\xi})$.

The above expressions are of course not applicable to non-Born interactions
such as contact interactions ($\Lambda\to\infty$, for which $a_{1},a_{2},a_{3}\to0$).
One can nevertheless easily generalise these expressions to the case
of non-Born interactions, by completing the first terms of the Born
expansion of $a$ (\emph{i.e.} performing the summation of ladder
diagrams in the language of perturbation theory). One arrives at:
\begin{align}
E_{P} & =\frac{2\pi\hbar^{2}}{m}n_{B}a\left(1+a/\tilde{\xi}+(a/\tilde{\xi})^{2}\right)\label{eq:PolaronSelfEnergy3}\\
\tilde{V}(\bm{r}) & =\frac{4\pi\hbar^{2}}{m}n_{B}a\Big(-(a/\tilde{\xi})f(r)+(a/\tilde{\xi})^{2}f(r)^{2}\Big)\label{eq:BipolaronPotential3}\\
\tilde{W}(\bm{x},\bm{y}) & =\frac{4\pi\hbar^{2}}{m}n_{B}\frac{a^{3}}{\tilde{\xi}^{2}}\Big(f(z)f(y)+f(x)f(y)+f(z)f(x)\Big)\label{eq:TripolaronPotential3}
\end{align}
It can be checked that Eq.~(\ref{eq:PolaronSelfEnergy3}) is indeed
consistent with the result $E_{P}=\frac{2\pi\hbar^{2}}{m}n_{B}a(1-\sqrt{2}a/\xi)^{-1}$
obtained by the coherent ansatz of Ref.~\cite{Shchadilova2016} and
to second order in $a$ with the diagrammatic result of Ref.~\cite{Christensen2015}.
In the rest of this paper, we will restrict our consideration to
the low-density limit $\Lambda^{-1},a\ll\xi$, which gives identical
results for both Born and non-Born interactions. In this situation,
the term proportional to $f(r)^{2}$ in the two-body potential of
Eqs.~(\ref{eq:BipolaronPotentialPerturbative}) and (\ref{eq:BipolaronPotential3})
becomes negligible, and one retrieves the well-known Yukawa attraction
mediated between two impurities by a bosonic medium~\cite{Bardeen1967,Pethick2002,Yu2012,Naidon2018a,Camacho-Guardian2018}.

\section{Mean-field theory}

An important observation is that the three-body potential, being of
third order in $a$, can be either attractive or repulsive depending
of the sign of $a$, whereas the two-body potential, being of second
order, is always attractive irrespective of the sign of $a$. In the
case of a repulsive interaction $a>0$, there is therefore a competition
between the induced two-body attraction and three-body repulsion.
In this situation, it is tempting to think to that this competition
could stabilise the impurities into a liquid state of a certain density~\cite{Bulgac2002}.
However, this is possible only if the impurities remain mixed with
the bosonic medium. On the contrary, the repulsion between the impurities
and the bosons tends to separate them, reducing the strength of the
mediated interactions.

To understand this interplay, let us consider the case of bosonic
impurities. So far we have obtained an essentially exact form of the
Hamiltonian in a state corresponding to the vacuum of Fröhlich quasiparticles.
The wave function for the impurities remains to be specified to obtain
a variational ansatz for the Hamiltonian. Let us choose the Hartree-Fock
ansatz in which all the $N_{I}$ impurities occupy the $\bm{k}=\bm{0}$
mode. In this state, the Hamiltonian of Eq.~(\ref{eq:PerturbedHamiltonian})
gives the following mean-field energy:
\begin{align}
E & =\frac{1}{2}g_{B}Vn_{B}^{2}+gn_{B}N_{I}+\frac{g_{I}+g_{\text{2}}}{2V}N_{I}^{2}+\frac{g_{\text{3}}}{6V^{2}}N_{I}^{3}\label{eq:Mean-field-energy1}
\end{align}
where $g_{B}=U_{B}(\bm{0})$, $g=U(\bm{0})$, $g_{I}=U_{I}(\bm{0})$,
$g_{2}=V(\bm{0})=-g^{2}/g_{B}$ and $g_{3}=W(\bm{0},\bm{0})=3g^{3}/(2g_{B}^{2}n_{B})$
represent the low-energy coupling constants of the system. In contrast,
the usual mean-field theory, on which previous theoretical works~\cite{Esry1997,Pu1998,Timmermans1998,Ao1998,Trippenbach2000,Pethick2002}
are based, is obtained by taking a Hartree-Fock ansatz for the original
Hamiltonian (\ref{eq:Hamiltonian}), resulting in Eq.~(\ref{eq:Mean-field-energy1})
with $g_{2}=g_{3}=0$. As we shall see, the mean-field energy of Eq.~(\ref{eq:Mean-field-energy1})
can give a lower variational upper bound of the exact ground-state
energy. 

It should be noted that within the Hartree-Fock ansatz, the coupling
constants $g_{2}$ and $g_{3}$ are obtained in the Born approximation.
However, the induced two- and three-body interactions $V$ and $W$
may or may not be perturbative, depending on the density of the bosons
$n_{B}$ and the mass $M$ of the impurities. The Born expansion
of $g_{2}$ and $g_{3}$ reads:
\begin{align}
g_{2} & =V(\bm{0})-\frac{1}{V}\sum_{\bm{p}}\frac{V(\bm{p})^{2}}{\frac{\hbar^{2}p^{2}}{M}}+\dots\nonumber \\
 & =-\frac{\pi\hbar^{2}}{m}\frac{a^{2}}{a_{B}}-\frac{\pi\hbar^{2}}{2m}\frac{M}{m}a^{4}\sqrt{\frac{n_{B}\pi}{a_{B}^{3}}}+\dots\label{eq:g2-result}\\
g_{3} & =W(\bm{0},\bm{0})-\frac{1}{V^{2}}\sum_{\bm{k},\bm{p}}\frac{W(\bm{k},\bm{p})^{2}}{\frac{\hbar^{2}}{M}(\bm{k}+\bm{p}/2)^{2}+\frac{3\hbar^{2}}{4M}p^{2}}+\dots\nonumber \\
 & =\frac{\pi\hbar^{2}}{m}\frac{3a^{3}}{4n_{B}a_{B}^{2}}-\frac{\hbar^{2}}{m}\frac{M}{m}\frac{a^{6}}{a_{B}^{2}}\underbrace{\lambda}_{17.5755}+\dots\label{eq:g3-result}
\end{align}

For the two-body and three-body potentials to satisfy the Born approximation,
it is thus required that 
\begin{equation}
\alpha_{\text{2}}\equiv\frac{M}{m}\frac{1}{2}\sqrt{\frac{\pi a^{4}n_{B}}{a_{B}}}\ll1\;\text{ and }\;\alpha_{\text{3}}\equiv\frac{4}{3}\lambda\frac{M}{m}n_{B}a^{3}\ll1.\label{eq:PerturbativeConditions}
\end{equation}
In addition to these requirements, the validity of the above mean-field
energy is limited by the two-body and three-body diluteness conditions,
\begin{equation}
\lambda_{\text{2}}\equiv n_{I}\left(L_{\text{2}}\right)^{3}\ll1\quad\text{ and }\quad\lambda_{\text{3}}\equiv n_{I}(L_{\text{3}})^{3}\ll1\label{eq:DilutenessCondition2B}
\end{equation}
where $n_{I}=N_{I}/V$ is the density of impurities, $L_{2}=(g_{I}+g_{2})/(4\pi\hbar^{2}/M)$
is the effective impurity two-body scattering length, and $L_{\text{3}}=[g_{3}/(4\pi\hbar^{2}/M)]^{1/4}$
is the three-body interaction length. In any case, however, the variational
nature of the mean-field energy Eq.~(\ref{eq:Mean-field-energy1})
ensures that it remains an upper bound of the exact ground-state energy.

The mean-field energy Eq.~(\ref{eq:Mean-field-energy1}) has been
written assuming that the whole system is homogeneous. This discards
the possibility of the impurities forming a separate phase. We thus
need to make a more general ansatz, in which the $N_{I}$ impurities
occupy a volume $V^{\prime}$ within the total volume $V$. Among
the total number of bosons $N_{B}$, a certain number $N_{B}^{\prime}$
may permeate in-between the impurities inside the volume $V^{\prime}$,
forming a density $n_{B}^{\prime}=N_{B}^{\prime}/V^{\prime}$. Let
us define the reduced energy $\tilde{E}=(E-\frac{1}{2}g_{B}Vn_{B}^{2})/(gn_{B}N_{I})$
as the energy difference between the total energy $E$ and the energy
of the pure homogeneous Bose gas of density $n_{B}$, normalised by
the interaction energy $gn_{B}N_{I}$. In the limit of large volume
$V$, the reduced energy of the completely segregated phase containing
a pure bubble of impurities ($n_{B}^{\prime}=0$) is simply:
\begin{equation}
\tilde{E}_{\text{pure}}=\frac{1}{2}v+\chi\frac{1}{2v}\label{eq:PureBubbleEnergy}
\end{equation}
where $v=\frac{g_{B}n_{B}V^{\prime}}{gN_{I}}$ is a reduced volume
and $\chi=\frac{g_{I}g_{B}}{g^{2}}$ is a parameter characterising
the relative strengths of interactions. One can easily find the minimum
$\tilde{E}_{\text{pure}}=\sqrt{\chi}$ at $v=\sqrt{\chi}$~\cite{Ao1998}.
On the other hand, in the mixed phase $n_{B}^{\prime}=fn_{B}$ where
there is inside the impurity volume $V^{\prime}$ a nonzero fraction
$f$ of the outside boson density, the reduced energy is given by:
\begin{equation}
\tilde{E}_{\text{mixed}}=\frac{1}{2}v(1-f)^{2}+f+\frac{1}{2v}(\chi-1)+\frac{1}{4v^{2}f}\label{eq:MixedBubbleEnergy}
\end{equation}

For $\chi\ge2$, minimising this energy with respect to $f$ and
$v$ gives $\tilde{E}_{\text{mixed}}=1$, with $f=1$ and $v\to\infty$,
corresponding to a completely mixed phase where impurities spread
into the condensate and form a uniform mixture. For $\chi<2$, however,
a non trivial solution of finite volume is found corresponding to
a bubble of polarons, \emph{i.e.} impurities mixed with bosons. This
bubble has a lower energy than a pure bubble of impurities because
of the induced two-body attraction. Moreover, its stability results
from the equilibrium between the induced three-body repulsion and
the pressure from the outside bosons. At the particular value $\chi=1$,
the induced two-body interaction exactly cancels the effects of the
direct repulsion between impurities, and the polarons form a Bose
gas interacting only with three-body interactions. Finally, at the
critical value $\chi_{c}=9/16=0.5625$, the energy of the polaron
bubble becomes equal to $\tilde{E}_{\text{pure}}$. At this point,
it is energetically more favourable for the polarons to collapse into
a bubble of impurities devoid of any bosons, about twice smaller than
the polaron bubble. This scenario is represented in Fig.~\ref{fig:Scenario}.

It should be noted that in this simple mean-field theory, the kinetic
energies of the bosons and impurities are neglected and densities
are assumed to be uniform inside and outside the volume $V^{\prime}$.
Taking into account the kinetic energies would smooth the densities
near the bubble surface and add a small surface tension that would
determine the bubble's geometry - a sphere in the present case of
a homogeneous Bose gas. Apart from these effects, the kinetic energies
may be neglected for a sufficiently large number of impurities. Indeed,
the contribution of the kinetic energies to $\tilde{E}$ is of the
order of $A_{B}=(N_{I}4\pi a/\xi)^{-1}$. Although $a/\xi$ is assumed
to be small, a mesoscopic number of impurities can easily make $A_{B}\ll1$. 

It is also worthwhile to point out that in this limit, the polaron
bubble cannot exist as a metastable state for $\chi\le\chi_{c}$.
Indeed, the activation energy for any local fluctuation in the polaron
bubble to nucleate a region free of bosons is solely due to the kinetic
energy cost of reducing the density of bosons to zero in that region,
which is proportional to $A_{B}$. For $A_{B}\ll1$, there is therefore
almost no energy cost, and the polaron bubble should quickly collapse
to an impurity bubble. The smallness of $A_{B}$ is also required
to guarantee that the range $\xi/\sqrt{f}$ of mediated interactions
remains smaller than the average spacing between the impurities. 

\begin{figure*}[t]
\begin{centering}
\includegraphics[width=12.5cm]{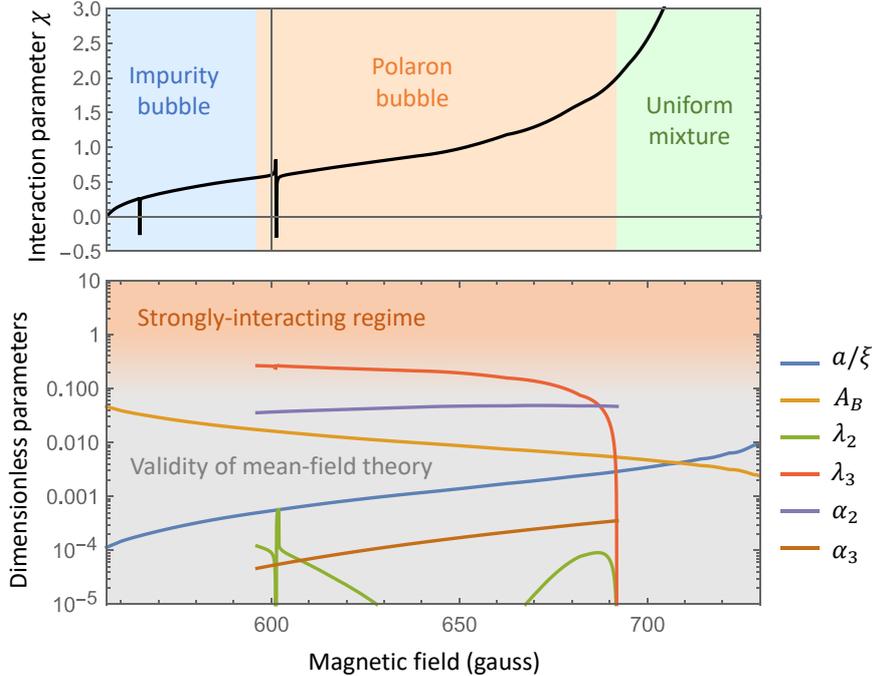}
\par\end{centering}
\caption{\label{fig:caesium-lithium}Expected phases of caesium-133 atoms immersed
in a condensate of lithium-7 atoms, as a function of applied magnetic
field. Top panel: interaction parameter $\chi=g_{I}g_{B}/g^{2}$.
Bottom panel: various dimensionless parameters {[}see Eqs.~(\ref{eq:PerturbativeConditions})
and (\ref{eq:DilutenessCondition2B}){]} characterizing the validity
of the Hartree-Fock ansatz for a dilute system, obtained for a total
number of caesium atoms $N_{I}\sim10^{5}$ and a density of lithium
atoms $n_{B}\sim10^{12}\text{ cm}^{-3}$. }
\end{figure*}

\section{Ultracold atomic mixtures}

Finally, let us consider how this physics could be observed with ultracold
atoms. The basic requirements are a mixture of heavy and light atoms
(for instance $m/M=0.05$) and a small ratio $a/\xi$ (for instance
0.0001). One should then specify the ratio $g_{B}/g$. A small ratio
ensures that the impurities remain a minority among the bosons (for
instance $g_{B}/g=0.2$, giving $n_{I}\approx0.2n_{B}^{\prime}$).
Near the three-body-dominated regime $\chi\approx1$ ($v\approx2$,$f\approx0.6$),
these values lead to the boson diluteness parameter $n_{B}a_{B}^{3}\approx4\times10^{-12}$,
the impurity 2-body and 3-body diluteness parameters $\lambda_{2}\approx5\times10^{-5}(\chi-1)^{3}$
and $\lambda_{3}\approx0.1$. The Born conditions of Eq.~(\ref{eq:PerturbativeConditions})
are also satisfied with $\alpha_{\text{2B}}\approx3\times10^{-3}$
and $\alpha_{\text{3B}}\approx2\times10^{-6}$. Taking a typical condensate
density $n_{B}=10^{15}\,\text{cm}^{-3}$, and the lightest boson available
(lithium-7), one finds $a_{B}\approx0.3a_{0}$, $a\approx3a_{0}$,
and $a_{I}\approx150a_{0}$, where $a_{0}$ designates the Bohr radius
$a_{0}\approx5.29\,\times\:10^{-11}$~m. These are moderate values
of atomic scattering lengths, for which atomic losses should remain
small. Requiring $A_{B}$ to be smaller than $0.01$ sets the number
of impurities to $\sim10^{6}$, which also corresponds to a realistic
number. The formation of a bubble could be observed by imaging the
impurity density, as in previous experiments, and the transition from
the impurity bubble to the polaron bubble could be evidenced by probing
the change of boson density inside the bubble, for instance by shining
a light causing inelastic collisions between bosons and impurities\footnote{This idea was suggested by Yoshiro Takahashi.}.

It should be pointed out that the energy difference $\Delta=(E_{\text{mixed}}-E_{\text{pure}})/N_{I}$
between the polaron bubble and the impurity bubble is only about 14
nK near $\chi\approx1$. Although it would seem that a very low temperature
$k_{B}T\ll\Delta$ is needed to observe the two phases distinctly,
previous experiments~\cite{Stenger1998,Papp2008} have shown that
such tiny interaction energies are enough to drive observable transitions
even at temperatures of the order of 100 nK. Interestingly, it is
possible to increase the energy difference by considering a higher
ratio $a/\xi$, for instance $a/\xi=$0.01, giving $\Delta\approx300$
nK. In this case, the polaron bubble becomes strongly interacting
with $\lambda_{2}\approx0.5(\chi-1)^{3}$ and $\lambda_{3}\approx1.0$,
along with $\alpha_{2}\approx0.35$ and $\alpha_{3}\approx0.02$.
Although the mean-field theory is not quantitative any more in this
regime, its variational nature indicates that a dense bubble of polarons
dominated by strong three-body interactions should exist near $\chi=1$.
These strong three-body interactions may indeed overcome the inelastic
three-body processes naturally occurring in ultracold atomic clouds.
To account for these inelastic processes, the three-body coupling
constant $g_{3}$ may be generalised to a complex number $(4\pi\hbar^{2}/M)\times(L_{3}^{4}-i\mathcal{L}_{3}^{4})$.
Since all scattering lengths $a_{B}$, $a$, and $L_{2}$ remain small
near $\chi=1$, the three-body loss rates $K=(h/M)\mathcal{L}_{3}^{4}$
should take an off-resonant value, typically $\sim10^{-28}\,\text{cm}^{6}\text{s}^{-1}$~\cite{Pethick2002},
corresponding to $\mathcal{L}_{3}\sim10\,\text{nm}$. On the other
hand, $L_{3}\sim100\text{ nm},$ which shows that elastic three-body
collisions may dominate inelastic ones in this regime.

As a concrete illustration, the expected phases of a mixture of lithium-7
and caesium-133 atoms (in their respective ground hyperfine states)
are shown in Fig.~\ref{fig:caesium-lithium}, as a function of applied
magnetic field. The figure is obtained for $N_{I}=10^{5}$ caesium
atoms, and a lithium density $n_{B}\sim10^{12}\text{ cm}^{-3}$. The
interaction parameter $\chi=g_{I}g_{B}/g^{2}$ is obtained from the
scattering lengths data for caesium-133~\cite{Berninger2013}, lithium-7~\cite{Pollack2009a},
and caesium-133 -- lithium-7~\cite{Naidon2019}. The polaron bubble
phase is expected in a 100~gauss-wide window. Even for the relatively
low density considered, the polaron bubble appears to be dense with
respect to the induced three-body interactions, with $\lambda_{\text{3}}$
approaching unity.

\section{Conclusion}

In summary, it has been shown that mixtures of bosons with repulsive
interactions may not only fully mix or fully separate, but also form
polaron bubbles. In contrast to usual dilute gases, these polaron
bubbles can be rather dense and dominated by three-body interactions.
These bubbles therefore provide a unique setting in the context of
ultracold atomic gases to study strongly-interacting gases with elastic
three-body interactions, a situation that has been so far out of experimental
reach.

\noindent \rule[0.5ex]{1\columnwidth}{1pt}

The author would like to thank Hiroyuki Tajima, Shimpei Endo, Ludovic
Pricoupenko, Munekazu Horikoshi, Tetsuo Hatsuda, and Yoshiro Takahashi
for useful discussions. The author is also grateful to Paul Julienne
for providing the caesium scattering length data. This work was supported
by the RIKEN Incentive Research Project and JSPS Grants-in-Aid for
Scientific Research on Innovative Areas (No. JP18H05407).


\end{document}